\documentclass[preprint,1p,a4paper,10pt,onecolumn]{elsarticle}
\biboptions{sort&compress}
\usepackage{amsmath}
\usepackage{amsfonts}
\usepackage{amssymb}
\usepackage{graphicx}
\usepackage{hyperref}
\PassOptionsToPackage{hyphens}{url}
\begin{document}
	\title{Parton distribution functions from scalar light-front parton gas model}
	\author[1,2,3]{Shaoyang Jia\corref{cor1}}
	\ead{sjia@iastate.edu}
	\cortext[cor1]{Corresponding author}
	\author[1]{James P. Vary\corref{cor2}}
	\ead{jvary@iastate.edu}
	\affiliation[1]{organization={Department of Physics and Astronomy, Iowa State University},
		city={Ames},
		postcode={Iowa 50011},
		country={U.\,S.\,A.}}
	\affiliation[2]{organization={Physics Division, Argonne National Laboratory},
		addressline={9700 South Cass Avenue},
		city={Lemont},
		postcode={Illinois 60439},
		country={U.\,S.\,A.}}
	\affiliation[3]{organization={Institute of Modern Physics, Chinese Academy of Sciences},
		addressline={509 Nanchang Road},
		city={Lanzhou},
		postcode={Gansu 730000},
		country={P.\,R.\,China}}
	\date{\today}
	\begin{abstract}
		We propose an application of a microcanonical ensemble with light-front kinematics to model the phase-space distribution of relativistic constituents of a bound state. These constituents denoted by partons are treated as classical spin-zero particles confined inside the bound state with inter-parton collisions as their only interaction. The microcanonical molecular dynamics ensemble is applied to obtain the phase-space distribution of such a thermodynamic system. We sample this phase-space distribution using Monte Carlo algorithms to obtain the parton distribution functions (PDFs) in scenarios with $3$, $4$, and $5$ identical partons. In addition PDFs when a selected number of massless partons are mixed with $3$ massive partons are also presented.
	\end{abstract}
	
	\maketitle
\section{Introduction\label{sc:intro}}
The parton distribution functions (PDFs) describe how the longitudinal momentum of hadrons is carried by quarks and gluons and are accessible, for example, through deep inelastic scattering experiments~\cite{CooperSarkar:1997jk}. Recent global fits of the nucleon and the pion PDFs are available in Refs.~\cite{Accardi:2016qay,Ball:2017nwa,Barry:2018ort,Hou:2019efy,Bailey:2020ooq,Barry:2021osv}. Calculating the PDF requires the nonperturbative solutions of quantum chromodynamics (QCD)~\cite{CTEQ:1993hwr}. The dependence of these PDFs on the factorization scale is given by the Dokshitzer--Gribov--Lipatov--Altarelli--Parisi equation~\cite{Dokshitzer:1977sg,Gribov:1972ri,Altarelli:1977zs,CTEQ:1993hwr}. Various nonperturbative approaches provide predictions for the PDFs at their corresponding scales~\cite{Steffens:1993pc,Steffens:1994hf,Diakonov:1996sr,Diakonov:1997vc,Mankiewicz:1996ep,Weigl:1996ii,Detmold:2001dv,Ball:2016spl,Chen:2017gck,Orginos:2017kos,Lin:2018qky,Liu:2018uuj,Izubuchi:2018srq,deTeramond:2018ecg,Karpie:2018zaz,Bednar:2018mtf}. While several statistical descriptions of the hadron PDFs for the nucleons and the pions are available~\cite{Mangano:1995qv,Bhalerao:1996fc,Bourrely:2001du,Bourrely:2015kla,2018arXiv180203153B}, in this article we propose a unique approach that relates the PDFs to the statistics of light-front dynamics. We work with the natural units where the speed of light in vacuum, the reduced Planck constant, and the Boltzmann constant are all $1$. We also adopt $g^{\mu\nu} = {\mathrm{diag}\{ 1, -1, -1, -1 \}}$ as the metric tensor of spacetime \cite{Peskin:1995ev}.

In the light-front quantization approach to solve QCD bound state problems, the structure of a hadron is specified by the light-front wavefunctions of its partons. Subsequently the PDFs can be calculated based on these wavefunctions as eigenvectors of the light-front Hamiltonian~\cite{Brodsky:1997de,Lan:2019vui,Lan:2019rba}. For example, the heavy sea quark components of the nucleons can be accessed with the light-front wavefunction beyond the leading Fock sector~\cite{Brodsky:1980pb,Brodsky:1981se}. In the basis light-front quantization (BLFQ) approach, the light-front wavefunction is solved from the eigenvalue problem of the light-front Hamiltonian in a basis-function representation~\cite{Vary:2009gt}. The wavefunctions for the valence quarks of mesons can be obtained within the BLFQ approach~\cite{Li:2015zda,Li:2017mlw,Tang:2018myz,Jia:2018ary,Lan:2019vui}. Calculations of the light-front wavefunctions beyond the leading Fock sector are also available, but require extra numerical efforts~\cite{Vary:2018pmv,Zhu:2024awq,Yu:2024mxo}. Simple model populations of Fock spaces have been considered without explicit interactions where basis-space enumeration is feasible~\cite{Vary:2009gt}. We here investigate model populations, \textit{i.e.} the density of microstates, that include the role of kinetic energy and conserved multi-particle kinematics in the system's phase-space distributions using the microcanonical ensemble.

PDFs calculable from the light-front wavefunctions also yield the probability of finding a parton carrying a specific amount of the total hadron momentum. Such an interpretation motivates the formulation of parton dynamics in terms of probability distributions instead of amplitudes. This is analogous to the phase-space formulation of non-relativistic quantum mechanics~\cite{Moyal:1949sk}. While angular-momentum dependent PDFs and certain limits of generalized parton distributions also yield probability interpretations~\cite{Jia:2012wf,Belitsky:2005qn}, their relations to the phase-space distribution of partons are beyond the scope of the model to be introduced in this article. 

In our initial investigation of the light-front parton gas model, the partons are treated as spin-zero classical particles confined inside an isolated bound state. These partons reach thermal equilibrium through elastic collisions with each other. The Hamiltonian of the thermal system is thus given by the light-front kinetic energy. The system has a fixed number of particles with conserved total light-front 3-momentum. With less than a dozen partons, the phase-space density of such a system is conveniently given by the microcanonical molecular dynamics ensemble~\cite{Lado:1981,ROMAN199653}. The PDFs are then obtained through the marginalization of the phase-space distributions specified by such an ensemble, during which both analytical integrals and the Gibbs sampling algorithm are applied~\cite{robert1999monte}. 

Statistical quantum field theories can be formulated in the grand-canonical ensemble with light-front quantization conditions~\cite{Raufeisen:2004dg,Beyer:2005rd,Strauss:2008zx,Raufeisen:2004ju,Raufeisen:2005cp}. However in our model based on the light-front Hamiltonian of classical particles, the microcanonical ensemble directly gives the joint probability distribution of the phase-space variables for all particles. Such a description is confirmed by the quantum-field calculation of BLFQ~\cite{Tang:2019gvn,Maris:2020wew,Tang:2020srr}. While for massless partons the end-point behavior of the longitudinal momentum-fraction distribution agrees with the quark counting rules~\cite{Ball:2016spl,deTeramond:2018ecg}. Our statistical description has the additional advantage of providing phase-space distributions as functions of light-front kinematic variables in terms of thermodynamic properties.

This article is organized as follows. Following this introduction, Sec.~\ref{sc:EVNP_ensemble} defines the light-front parton gas model with its phase-space distribution given by the microcanonical molecular dynamics ensemble. We present the single-particle longitudinal momentum-fraction distribution in Sec.~\ref{sc:margin_rho} as our modeling of the PDF. The summary and concluding remarks are given in Sec.~\ref{sc:summary}. 
\section{The microcanonical molecular dynamics ensemble description of the light-front parton gas\label{sc:EVNP_ensemble}}
The light-front quantization condition of a quantum field theory is specified at a fixed light-front time ${x^+=x^0+x^3}$. The longitudinal spatial coordinate is given by $x^-=x^0-x^3$ orthogonal to $x^+$. The conjugate momenta of $x^+$ and $x^-$ are $p^-=p^0-p^3$ and $p^+=p^0+p^3$, respectively~\cite{Brodsky:1997de}. 
In the initial investigation of the light-front parton gas model, we ignore both statistics and dynamics of particle spin. We further ignore quantum field theory effects and arrive at a parton model whose dynamics is given by classical mechanics. Particle creation and annihilation are also disabled, resulting in systems with a fixed number of particles. 

Let us consider the parton gas system whose Hamiltonian consists of the light-front kinetic energy for non-interacting scalar partons:
\begin{equation}
	P^-(\mathbf{q},\mathbf{p}) = \sum_{j=1}^{N} \left( \overrightarrow{p}^{\perp 2}_{j}+m^2_{j} \right ) / p_j^+.\label{eq:def_P-_LFPG}
\end{equation} 
Here $\overrightarrow{p}^\perp_j$ and $p_j^+$ are the transverse momentum and the longitudinal momentum of the $j$-th parton. The vector $\mathbf{p}$ represents all of the $3N$ momentum variables $\{ p_j^+,\, {\overrightarrow{p}^\perp_j} \}$. Notice that all $p_j^+$ are positive definite for massive particles, because the point of ${p_j^+ = 0}$ cannot be reached when the total energy is finite. The limit of ${p_j^+ \rightarrow 0}$ is a kinematic singularity when the particles are massless, but measures $0$ after integrations in the phase space. Although there is no coordinate-space interaction in Eq.~\eqref{eq:def_P-_LFPG}, we keep the vector $\mathbf{q}$ to represent the light-front coordinates of all partons. The mass of the $j$-th parton is given by $m_j$, allowing the system under consideration to consist particles of different masses.

We are interested in the structure of an isolated bound state. Therefore the system of partons is not in thermal contact with a reservoir, nor does it exchange particles with another source. We also assume that partons reach thermal equilibrium before the bound state decays. The phase-space distribution of the partons is consequently given by the microcanonical ensemble. 

Since there is no interaction in the coordinate space, the confinement of partons inside the bound state is naively implemented by the finite supports of the coordinate-space integrals. With appropriate boundary conditions, the classical system whose dynamics is specified by Eq.~\eqref{eq:def_P-_LFPG} conserves the total light-front $3$-momentum. In this case the microcanonical ensemble is further constrained into the microcanonical molecular dynamics ensemble~\cite{Lado:1981,ROMAN199653}, with its phase-space distribution given by
\begin{equation}
	\rho\left(E,V,N,\mathbf{P};\mathbf{q},\mathbf{p}\right) = \dfrac{ \delta\left(E-P^-\left(\mathbf{q},\mathbf{p}\right)\right) }{ \Omega(E,V,N,\mathbf{P}) } \, \delta \bigg( \mathbf{P}-\sum_{j=1}^{N}\mathbf{p}_j \bigg).\label{eq:EVNP_ensemble_rho}
\end{equation}
This phase-space distribution corresponds to the ergodic averaging of molecular dynamics simulations. The partition function $\Omega(E,V,N,\mathbf{P})$ normalizes the phase-space distribution through its definition:
\begin{equation}
	\Omega(E,V,N,\mathbf{P}) = \! \int d^{3N}\mathbf{q} \int d^{3N} \mathbf{p} 
	\, \rho (E,V,N,\mathbf{P};\mathbf{q},\mathbf{p} ).\label{eq:def_partition_function}
\end{equation} 
Corrections from the phase-space volume may be required to calculate the correct entropy of the system from the partition function~\cite{ROMAN199653}. The first Dirac $\delta$-function in Eq.~\eqref{eq:EVNP_ensemble_rho} restricts the system to a specific total energy $E$. The second $\delta$-function defined as 
\begin{equation}
	\delta \bigg( \mathbf{P}-\sum_{j=1}^{N}\mathbf{p}_j \bigg) = \delta \bigg(P^+ -\sum_{j=1}^{N}p_j^+ \bigg) \, \delta \bigg( \overrightarrow{P}^\perp -\sum_{j=1}^{N}\overrightarrow{p}^\perp_j \bigg) \label{eq:def_P_delta}
\end{equation}
ensures the conservation of the total light-front $3$-momentum of the system. 

The microcanonical molecular dynamics ensemble having fixed energy $E$, volume $V$, particle number $N$ and total momentum $\mathbf{P}$, is therefore also called the EVN\textbf{P} ensemble. Notice that due to momentum conservation indicated by Eq.~\eqref{eq:def_P_delta}, the number of degrees of freedom for the momentum variables is reduced from $3N-1$ in the microcanonical ensemble to $3N-4$ in the EVN\textbf{P} ensemble. Based on the partition function $\Omega(E,V,N,\mathbf{P})$ given by Eq.~\eqref{eq:def_partition_function}, one can define entropy, temperature, pressure, chemical potential, and other thermodynamic quantities~\cite{ROMAN199653}.

After specifying the light-front Hamiltonian $P^-(\mathbf{q},\mathbf{p})$, the single-particle momentum distribution of the $\alpha$-th parton can be defined as the marginal distribution of the phase-space distribution in Eq.~\eqref{eq:EVNP_ensemble_rho}:
\begin{equation}
	\omega_{\alpha} ( p^+,\overrightarrow{p}^\perp ) = \int d^{3N}\mathbf{q}\, \bigg( \prod_{a=1,a\neq \alpha}^{N}\int d\,\mathbf{p}_a \bigg) \, \rho(E,V,N,\mathbf{P};\mathbf{q},\mathbf{p}).\label{eq:def_single_particle_momentum_dist}
\end{equation}
Particles whose mass is identical to $m_{\alpha}$ share the same distribution ${ \omega_{\alpha} ( p^+,\overrightarrow{p}^\perp ) }$. One could further marginalize Eq.~\eqref{eq:def_single_particle_momentum_dist}, leaving only the momentum components of interest. 

We then proceed to simplify the phase-space distribution $\rho(\mathbf{q},\mathbf{p})$ with the light-front kinetic energy written in Eq.~\eqref{eq:def_P-_LFPG}. For a system with fixed total light-front $3$-momentum, it becomes convenient to define longitudinal momentum fractions $x_j$ and relative transverse momenta $\overrightarrow{\kappa}^{\perp}_{j}$. They are related to the single-particle momentum $(p^+_j,\,\overrightarrow{p}^\perp_j)$ and system total momentum $(P^+,\,\overrightarrow{P}^\perp)$ through
\begin{equation}
	\begin{cases}
		x_j=p_j^+/P^+\\[1mm]
		\overrightarrow{\kappa}_j^\perp=\overrightarrow{p}_j^\perp-x_j\overrightarrow{P}^\perp
	\end{cases},\label{eq:def_relative_momenta}
\end{equation}
with the total momenta defined as $P^+=\sum_{j=1}^{N}p_j^+$ and $\overrightarrow{P}^\perp =\sum_{j=1}^{N}\overrightarrow{p}_j^\perp$. The Jacobian due to Eq.~\eqref{eq:def_relative_momenta} is given by
\begin{equation}
	d^{3N}\mathbf{p} = (P^+)^N d^N x\, d^{2N}\mathbf{\kappa}^\perp = (P^+)^N \prod_{j=1}^{N}dx_j\, d\overrightarrow{\kappa}_j^\perp.
\end{equation}
In terms of these relative momentum variables, the Hamiltonian in Eq.~\eqref{eq:def_P-_LFPG} can be written as 
\begin{equation}
	P^-(\mathbf{q},\mathbf{p}) = \dfrac{1}{P^+} \bigg( \overrightarrow{P}^{\perp 2} + \sum_{j=1}^{N}\dfrac{\overrightarrow{\kappa}^{\perp 2}_j+m^2_j}{x_j} \bigg).\label{eq:LFPG_P-_relative_momenta}
\end{equation}
Meanwhile Eq.~\eqref{eq:def_P_delta} that ensures the conservation of the light-front $3$-momentum becomes
\begin{equation}
	\delta \bigg( \mathbf{P}-\sum_{j=1}^{N}\mathbf{p}_j \bigg) = \dfrac{1}{P^+}\delta \bigg( 1-\sum_{j=1}^{N}x_j \bigg) \, \delta \bigg( \sum_{j=1}^{N}\overrightarrow{\kappa}^\perp_j \bigg).\label{eq:P_delta_relative_momenta}
\end{equation}
Substituting Eqs.~\eqref{eq:LFPG_P-_relative_momenta} and \eqref{eq:P_delta_relative_momenta} into Eq.~\eqref{eq:EVNP_ensemble_rho} produces
\begin{align}
	& \rho\left(E,V,N,\mathbf{P};\mathbf{q},\mathbf{p}\right) = \dfrac{1}{\Omega(E,V,N,\mathbf{P})}\, \delta \bigg( 1-\sum_{j=1}^{N}x_j \bigg) \delta \bigg( \sum_{j=1}^{N}\overrightarrow{\kappa}^\perp_j \bigg) \nonumber\\
	& \times \delta \bigg( P^+E-\overrightarrow{P}^{\perp 2} - \sum_{j=1}^{N}\dfrac{\overrightarrow{\kappa}^\perp_j+m^2_j}{x_j} \bigg). \label{eq:rho_LFPG_ori}
\end{align}
Equation~\eqref{eq:rho_LFPG_ori} is the phase-space distribution of the light-front parton gas model written in the longitudinal momentum fractions and the relative transverse momenta. This phase-space distribution follows from full relativistic kinematics of light-front dynamics, therefore has the desirable properties of boost invariance in the longitudinal and in the transverse directions.

For numerical evaluations of the phase-space distribution, the $\delta$-functions in Eq.~\eqref{eq:rho_LFPG_ori} can be eliminated after marginalizing a subset of the phase-space variables. Explicitly in order to eliminate the first $2$ $\delta$-functions of Eq.~\eqref{eq:rho_LFPG_ori}, we integrate with respect to coordinates $\mathbf{q}$ and with respect to the light-front $3$-momenta of the $N$-th parton such that
\begin{align}
	& \int d^{3N}\mathbf{q} \int dp_N^+ \int d\overrightarrow{p}^\perp_N\, \rho (\mathbf{q},\mathbf{p}) \nonumber\\
	& = \dfrac{V^{3N}P^+}{\Omega(E,V,N,\mathbf{P})} \, \delta \bigg( u-\sum_{j=1}^{N-1}\dfrac{ \overrightarrow{\kappa}^{\perp 2}_j+m^2_j }{ x_j } - \dfrac{ ( \sum_{j=1}^{N-1}\overrightarrow{\kappa}^\perp _j )^2 + m^2_{N} }{ 1-\sum_{j=1}^{N-1}x_j } \bigg),\label{eq:itg_rho_step_1}
\end{align}
with 
\begin{equation}
	u=P^+E-\overrightarrow{P}^{\perp 2} \label{eq:def_u}
\end{equation}
being the thermal energy available to the light-front kinematics of relative parton motion. In the absence of coordinate-space interactions, the coordinate-space integrals are trivially related to the volume of the bound state by $V= (1/2) \int dx^-\int d\overrightarrow{x}^\perp$. The remaining $\delta$-function in Eq.~\eqref{eq:itg_rho_step_1} corresponds to roots of a quadratic function of $x_{N-1}$. Explicitly after simplifying this $\delta$-function we obtain
\begin{align}
	& \int d^{3N}\mathbf{q} \int dp_N^+ \int d\overrightarrow{p}^\perp_N \int dp_{N-1}^+\, \rho(\mathbf{q},\mathbf{p}) = \dfrac{ V^{3N}(P^+)^2 }{ \Omega(E,V,N,\mathbf{P}) } \nonumber\\
	& \times \int_{0}^{1} dx_{N-1} \theta( \eta^2 - 4\alpha_{N-1}\beta_{N-2} ) \, \dfrac{ \delta\left(x_{N-1}-x_+\right)+\delta\left(x_{N-1}-x_-\right) }{ \bigg\vert \dfrac{ \overrightarrow{\kappa}^{\perp 2}_{N-1} + m^2_{N-1} }{ x_{N-1}^{2} } -  \dfrac{ (\sum_{j=1}^{N-1} \overrightarrow{\kappa}^{\perp}_j )^2 + m^2_{N} }{ \beta_{N-1}^{2} } \bigg\vert },\label{eq:reduction_int_rho}
\end{align}
where the Heaviside step function ensures the relevant discriminant to be non-negative. Here for the notational convenience we have defined dimensionless quantities $\alpha_j$ and $\beta_j$ according to 
\begin{subequations}\label{ea:def_alpha_beta}
	\begin{equation}
		\alpha_j = \dfrac{\overrightarrow{\kappa}^{\perp 2}_j+m^2_j}{u-\sum_{i=1}^{j} ( \overrightarrow{\kappa}^{\perp 2}_i+m^2_i ) / x_i},
	\end{equation}
	and
	\begin{equation}
		\beta_j = 1-\sum_{i=1}^{j}x_i.
	\end{equation}
\end{subequations}
Quantities $x_{\pm}$, $\eta$, and $\zeta$ in Eq.~\eqref{eq:reduction_int_rho} are subsequently given by
\begin{subequations}\label{eq:def_eta_zeta}
	\begin{align}
	x_{\pm} & =(\eta\pm \zeta)/2,\\
	\eta & =\beta_{N-2}+\alpha_{N-1}-\alpha_{N},\\
	\zeta & =\sqrt{\eta^2-4\alpha_{N-1}\beta_{N-2}}.
	\end{align}
\end{subequations}
Equation~\eqref{eq:reduction_int_rho} is the starting point for the marginalization of $\rho(\mathbf{q},\mathbf{p})$ using quadrature. 
\section{Marginalization of the phase-space distribution\label{sc:margin_rho}}
\subsection{Analytical reduction of transverse momentum integrals\label{ss:red_energy_delta_analyic}}
In the scenario where only distributions of the longitudinal momentum fractions are of interest, the marginalization of the phase-space distribution by integrals with respect to $\mathbf{p}^\perp$ can be evaluated analytically. Let us start with the definition of the joint longitudinal momentum-fraction distribution of all partons: 
\begin{equation}
	\omega(E,V,N,\mathbf{P};\mathbf{x}) = (P^+)^N\int d^{2N}\mathbf{p}^\perp \int d^{3N}\mathbf{q} \, \rho(E,V,N,\mathbf{P};\mathbf{q},\mathbf{p}).\label{eq:def_omega_joint_x}
\end{equation}
Recall the phase-space distribution ${\rho(E,V,N,\mathbf{P};\mathbf{q},\mathbf{p})}$ is given by Eq.~\eqref{eq:rho_LFPG_ori} in the light-front parton gas model. In such a scenario Eq.~\eqref{eq:def_omega_joint_x} becomes 
\begin{align}
	& \omega(E,V,N,\mathbf{P};\mathbf{x}) = \dfrac{(P^+V)^N}{\Omega(E,V,N,\mathbf{P})} \delta \bigg( 1-\sum_{j=1}^{N}x_j \bigg) \nonumber\\
	& \times \int d^{2N} \kappa^{\perp} \,\delta \bigg( \sum_{j=1}^{N}\overrightarrow{\kappa}^\perp_j \bigg) \, \delta \bigg( u-\sum_{j=1}^{N}\dfrac{\overrightarrow{\kappa}^{\perp 2}_{j}+m^2_j}{x_j} \bigg).\label{eq:omega_joint_x_LFPQ}
\end{align}
We then consider the following substitution of integral variables:
\begin{equation}
	\overrightarrow{\kappa}^\perp_j=\sqrt{x_j}\overrightarrow{l}^\perp_j.\label{eq:kappa_to_q_perp}
\end{equation}
The change in the integral measure due to Eq.~\eqref{eq:kappa_to_q_perp} is 
\begin{equation}
	\int d^{2N}\kappa^\perp= \bigg( \prod_{j=1}^{N}x_j \bigg) \int d^{2N}l^\perp.
\end{equation}
Equation~\eqref{eq:omega_joint_x_LFPQ} then becomes
\begin{align}
	& \omega(E,V,N,\mathbf{P};\mathbf{x}) = \dfrac{ (P^+V)^N ( \prod_{j=1}^{N}x_j ) }{ \Omega(E,V,N,\mathbf{P}) } \delta \Big( 1-\sum_{j=1}^{N}x_j \Big) \nonumber\\
	& \times \int d^{2N} l^{\perp} \delta \Big( \tilde{u}(\mathbf{x})-\sum_{j=1}^{N}\overrightarrow{l}^{\perp 2}_j \Big) \, \delta \Big( \sum_{j=1}^{N}\sqrt{x_j}\overrightarrow{l}^\perp_j \Big),\label{eq:omega_joint_x_LFPQ_itg_l}
\end{align}
with 
\begin{equation}
	\tilde{u}(\mathbf{x})=u-\sum_{i=1}^{N} m^2_j / x_j\label{eq:def_u_tilde}
\end{equation}
and the available thermal energy $u$ given by Eq.~\eqref{eq:def_u}. If we define $\xi_j=\sqrt{x_j}$ as a vector in $N$-dimensional Euclidean space, the third $\delta$-function in Eq.~\eqref{eq:omega_joint_x_LFPQ_itg_l} can be viewed as that of inner products. Notice that $\xi_j$ is a unit vector due to the first $\delta$-function. Transverse integrals in Eq.~\eqref{eq:omega_joint_x_LFPQ_itg_l} can subsequently be calculated exactly in hyperspherical coordinates. As a result we obtain
\begin{subequations}\label{eq:def_joint_longi_dist}
\begin{equation}
	\omega(E,V,N,\mathbf{P};\mathbf{x}) = \dfrac{1}{\Phi(E,V,N,\mathbf{P})} \bigg( \prod_{j=1}^{N}x_j \bigg) \, \delta \bigg( 1-\sum_{j=1}^{N}x_j \bigg) \, [\tilde{u}(\mathbf{x})]^{N-2}\, \theta(\tilde{u}(\mathbf{x})),\label{eq:omega_joint_x}
\end{equation}
with $\Phi(E,V,N,\mathbf{P})$ defined as
\begin{equation}
	\Phi(E,V,N,\mathbf{P}) = \int d^Nx \, \bigg( \prod_{j=1}^{N}x_j \bigg) \, \delta \bigg( 1-\sum_{j=1}^{N}x_j \bigg) [\tilde{u}(\mathbf{x})]^{N-2}\,\theta(\tilde{u}(\mathbf{x}))
\end{equation}
\end{subequations}
ensuring the unit normalization of ${\omega(E,V,N,\mathbf{P};\mathbf{x})}$. Details in calculating the transverse integrals to derive Eq.~\eqref{eq:def_joint_longi_dist} are given in \ref{sc:transverse_itg}. Equation~\eqref{eq:omega_joint_x} is the analytical expression for the joint longitudinal momentum-fraction distribution of the light-front parton gas model. Since we do not discuss the entropy of the parton gas system, the relation between the normalization ${\Phi(E,V,N,\mathbf{P})}$ and the partition function ${\Omega(E,V,N,\mathbf{P})}$ is omitted. 
\subsection{Single-parton longitudinal momentum-fraction distribution}
In the light-front parton gas model, the probability of finding a parton carrying a given longitudinal momentum fraction can be obtained by marginalizing the phase-space distribution in Eq.~\eqref{eq:rho_LFPG_ori}. Such single-particle longitudinal moment-fraction distribution of the $i$-th parton is explicitly defined as
\begin{equation}
	\omega_{x,i}(E,V,N,\mathbf{P};x_i) = (P^+)^N \int d\overrightarrow{\kappa}^\perp_1 \Big( \prod_{j=1,\,j\neq i}^{N} \int dx_j \int d\overrightarrow{\kappa}_j^\perp \Big) \int d^{3N}\mathbf{q}\, \rho(E,V,N,\mathbf{P};\mathbf{q},\mathbf{p}).\label{eq:def_omega_x}
\end{equation}
For particles of identical mass, it does not matter which longitudinal momentum fraction $x_j$ is omitted by the integrals in Eq.~\eqref{eq:def_omega_x}. Equation~\eqref{eq:def_omega_x} can therefore be interpreted as the definition of the PDF for all particles of mass $m_i$ in our model. Assuming a total number of $M$ particles are identical, grouping them together can result in $m_j = m_1$ for $2 \leq j \leq M$. We subsequently choose $x_1$ as the default variable of their PDFs. When all particles are identical, we further omit the particle label in subscripts on the left-hand side of Eq.~\eqref{eq:def_omega_x}.

Specifically for a system of $2$ partons both of the same mass $m$, only the relative transverse momentum needs to be integrated in Eq.~\eqref{eq:def_omega_x}. After completing these integrals we have 
\begin{align}
	& \omega_x(E,V,2,\mathbf{P};x) = \dfrac{ 6\, x\, (1-x)\, \theta ( u-4m^2) }{ \left( 1 + 2m^2/u \right) \sqrt{1-4m^2/u}} \theta \left( x- 1/2 + \sqrt{ 1/4 - m^2/u } \right) \nonumber\\
	& \times \theta \left( 1/2 + \sqrt{ 1/4 - m^2/u } - x \right), \label{eq:omega_x_N2}
\end{align}
where the dependence on $E$ and $\mathbf{P}$ comes through the available thermal energy $u$ defined in Eq.~\eqref{eq:def_u}. The distribution $\omega_x(x)$ in Eq.~\eqref{eq:omega_x_N2} is independent of the volume because the coordinate-space integrals factor out, which is true in the absence of coordinate-space interactions. The $\theta$-functions in Eq.~\eqref{eq:omega_x_N2} reflect that the Hamiltonian defined by Eq.~\eqref{eq:LFPG_P-_relative_momenta} is positive semi-definite. Since the minimum contribution to the Hamiltonian $P^-(\mathbf{q},\mathbf{p})$ from the transverse momentum $\overrightarrow{\kappa}^\perp$ is zero, the available thermal energy $u$ needs to be greater than the longitudinal part of the kinetic energy. In the limit of ${u\rightarrow 4 \, m^2}$ Eq.~\eqref{eq:omega_x_N2} becomes ${\omega_x(x)=\delta(x-1/2)}$, which indicates no relative parton motion in such a limit. While in the limit of ${u \gg 4 \, m^2}$, the function in Eq.~\eqref{eq:omega_x_N2} is reduced to ${\omega_x(x)=6\,x\,(1-x)}$. This corresponds to either massless partons or extremely high thermal energy.

Aside from the marginalization of the phase-space distribution $\rho(\mathbf{q},\mathbf{p})$ directly using Eq.~\eqref{eq:def_omega_x}, the longitudinal momentum-fraction distribution for a single parton can also be calculated by marginalizing the joint distribution given by Eq.~\eqref{eq:omega_joint_x}. This single-particle $x$-distribution of the $i$-th particle is explicitly defined as 
\begin{equation}
	\omega_{x,i}(E,V,N,\mathbf{P};x_i) = \bigg( \prod_{j=1,\,j\neq i}^{N}\int dx_j \bigg)\, \omega(E,V,N,\mathbf{P}; \mathbf{x}). \label{eq:omega_x_from_joint}
\end{equation}
Such an alternative definition is in agreement with Eq.~\eqref{eq:def_omega_x}. 

In the limit where all particles are massless, the $\tilde{u}(\mathbf{x})$ defined by Eq.~\eqref{eq:def_u_tilde} is independent of the longitudinal momentum fraction $x$, reducing to $u$ in Eq.~\eqref{eq:def_u}. In this limit there is no scale for the relative motion of the partons, resulting in $\omega_x(x)$ only depending on the particle number $N$. Using induction one can demonstrate that 
\begin{equation}
	\omega_x(E,V,N,\mathbf{P};x) = (2N-2)(2N-1) \, x \, (1-x)^{2N-3} \, \theta \big( P^+E-\overrightarrow{P}^{\perp 2} \big).\label{eq:omega_x_N_massless}
\end{equation}
Equation~\eqref{eq:omega_x_N_massless} gives the $\omega_x(x)$ of the massless light-front parton gas model. The large-$x$ behavior given by Eq.~\eqref{eq:omega_x_N_massless} is in agreement with the quark counting rules~\cite{Ball:2016spl,deTeramond:2018ecg}. 

With $3$ identical massive partons, we calculate the single-particle longitudinal momentum-fraction distribution by marginalizing Eq.~\eqref{eq:omega_joint_x}. Specifically we apply the variable transformations ${\xi=x_2+x_3}$ and ${\lambda=x_2 x_3}$ such that
\begin{equation*}
	\int_{0}^{1-x_1}dx_2 \int_{0}^{1-x_1-x_2}dx_3\, \theta(\tilde{u}(\mathbf{x})) = \int_{0}^{1-x_1}d\xi\int_{ x_1 (1-x_1) / (ux_1-1) }^{ (1-x_1)^2 / 4 }d\lambda\, \dfrac{\theta(\xi^2-4\lambda)}{\sqrt{\xi^2-4\lambda}}.
\end{equation*}
In the units where $m=1$, the single-particle $x$-distribution becomes
\begin{equation}
	\omega_x(x) = \dfrac{ \left[(1-x)(x_+-x)(x-x_-) \right]^{3/2} }{ \phi(u)\, \sqrt{ux-1} } \theta(u-9) \, \theta(x-x_-)\, \theta(x_+-x) \label{eq:omega_x_N3_massive}
\end{equation}
with 
\begin{equation}
	x_\pm = \left[ u-3 \pm \sqrt{(u-9)(u-1)} \right] /(2u).
\end{equation}
The normalization $\phi(u)$ is defined as
\begin{align}
	\phi(u) = \int_{x_-}^{x_+} dx\, \dfrac{\left[(1-x)(x_+-x)(x-x_-) \right]^{3/2}}{\sqrt{ux-1}}. 
\end{align}
Again the $\theta$-functions in Eq.~\eqref{eq:omega_x_N3_massive} reflect the positive semi-definiteness of the Hamiltonian. One can verify Eq.~\eqref{eq:omega_x_N3_massive} is reduced to Eq.~\eqref{eq:omega_x_N_massless} for ${N=3}$ in the limit of ${u\rightarrow+\infty}$.
\subsection{Quadrature marginalization of the phase-space distribution}
\begin{figure*}
	\centering
	\includegraphics[width=\linewidth]{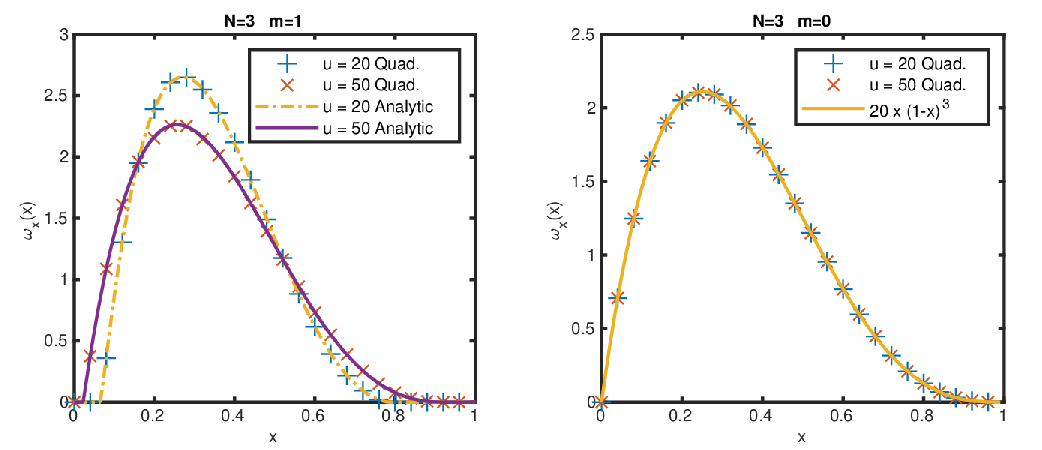}
	\caption{The single-particle longitudinal momentum-fraction distribution from the light-front parton gas model with $3$ partons of the same mass. On the left panel, the partons are massive with the available thermal energy $u$ measured in units of $m^2$. The blue stars and red crosses correspond to quadratures of Eq.~\eqref{eq:omega_x_marginalized} at $u = 20\,m^2$ and $u = 50\,m^2$ respectively. The yellow dot-dashed line and the purple solid line are the results from the analytical expression given by Eq.~\eqref{eq:omega_x_N3_massive} at the corresponding values of $u$. On the right panel, the partons are massless, where the units of available thermal energy is the only scale. Discrete points are results from quadrature using Eq.~\eqref{eq:omega_x_marginalized} at $u = 20$ and $50$. The $\omega_x(x)$ obtained from quadrature are independent of $u$. They agree with the analytical result ${20\,x(1-x)^3}$ within numerical errors.}
	\label{fig:quad_N3}
\end{figure*}
When all partons are of the same non-vanishing mass, this mass sets the scale of the phase-space distribution. Therefore we choose ${m=1}$ setting the default units of measure. The marginalized distribution $\omega_x(x)$ with ${N=2}$ is already given by Eq.~\eqref{eq:omega_x_N2}. When we do not impose the condition that particles have the same mass, in the case of ${N\geq 3}$ we substitute Eq.~\eqref{eq:reduction_int_rho} into Eq.~\eqref{eq:def_omega_x} and obtain
\begin{align}
	& \omega_{x,i}(E,V,N,\mathbf{P};x_i) = \dfrac{V^{3N}(P^+)^N}{\Omega(E,V,N,\mathbf{P})} \bigg( \prod_{j=1}^{N-1} \int d\overrightarrow{\kappa}_j^\perp \bigg) \bigg( \prod_{j=1,\,j\neq i}^{N-1}\int dx_j \bigg)_{\mathbf{x}\in D_i(\mathbf{x})}\nonumber\\
	& \times \dfrac{ \delta (x_{N-1}-x_+) + \delta (x_{N-1}-x_-) }{ \bigg\vert \dfrac{ \overrightarrow{\kappa}^{\perp 2}_{N-1}+m^2_{N-1} }{ x_{N-1}^{2} } - \dfrac{ ( \sum_{j=1}^{N-1}\overrightarrow{\kappa}^{\perp}_j )^2 + m^2_{N} }{ \beta_{N-1}^{2} } \bigg\vert } \, \theta(\eta^2-4\alpha_{N-1}\beta_{N-2}),\label{eq:omega_x_marginalized}
\end{align}
with $x_\pm$, $\eta$, $\alpha_j$, and $\beta_j$ defined by Eqs.~\eqref{eq:def_eta_zeta} and \eqref{ea:def_alpha_beta}. The subscript condition of the longitudinal momentum-fraction integrals is explicitly given by 
\begin{equation}
	\mathbf{x}\in D_i(\mathbf{x}) = \bigg\{ x_j~(~\mathrm{for}~1 \leq j\leq N-1~\mathrm{and}~j\neq i~)\, \bigg\vert\, 
	x_j\in (0,1) ~ \mathrm{and} \, \sum_{j=1,\,j\neq i}^{N-1}x_j < 1-x_i \bigg\},\label{eq:subscript_cond_X}
\end{equation}
which ensures that all longitudinal momenta $p_j^+$ are positive definite as required by the light-front kinematics.

Equation~\eqref{eq:omega_x_marginalized} is the expression of the single-particle longitudinal momentum-fraction distribution from the light-front parton gas model alternative to Eq.~\eqref{eq:omega_x_from_joint}. After eliminating the $x_{N-1}$ integral using the $\delta$-functions in Eq.~\eqref{eq:omega_x_marginalized}, the number of dimensions for remaining integrals is $3N-5$. This agrees with the number of momentum-space dimensions $3N$ subtracting the conservation of light-front $3$-momentum, $1$ conservation of energy, and $1$ remaining variable after marginalization. 

We then compute Eq.~\eqref{eq:omega_x_marginalized} with ${N=3}$ identical particles numerically using quadrature in the units where $m=1$. Since each transverse kinematic term in the Hamiltonian given by Eq.~\eqref{eq:def_P-_LFPG} is positive semi-definite, for any component of the transverse momentum being integrated we have a natural cutoff
\begin{equation}
	\vert \kappa_{j,\,x/y} \vert \leq \max \big( \sqrt{ x_j u - m^2 } \big) = \sqrt{u-m^2} = \kappa_{\mathrm{max}},
\end{equation}
where $\kappa_{j,\,x/y}$ stands for either the $x$- or the $y$-component of $\overrightarrow{\kappa}^\perp_j$. We sample transverse momentum components ranging from $-\kappa_{\mathrm{max}}$ to $\kappa_{\mathrm{max}}$ by the Simpson's rule after the variable transformation ${\kappa_j = \kappa_{\mathrm{max}} \,\mathrm{arctanh}\,(\phi_j)}$. For the numerical result to be accurate within $1\%$ difference from the analytical result, we allocate $99$ equally spaced quadrature points for each component of the transformed variable $\phi_j$ in the computation with massive partons. While the same number is increased to $199$ for massless partons. The resulting distributions $\omega_x(x)$ at ${u=20\,m^2}$ and ${u=50\,m^2}$ are presented on the left panel of Fig.~\ref{fig:quad_N3}. For the massless partons, the numerical result for $\omega_x(x)$ is shown on the right panel of Fig.~\ref{fig:quad_N3}. Both numerical results are in agreement with the analytical expressions given by Eqs.~\eqref{eq:omega_x_N_massless} and \eqref{eq:omega_x_N3_massive}. Therefore applying quadrature in the specific case of ${N=3}$, we have tested the analytical reduction of the transverse integrals in Subsection~\ref{ss:red_energy_delta_analyic}. In the case of ${N>3}$, the support of $x_j$ integrals in Eq.~\eqref{eq:omega_x_marginalized} is specified by Eq.~\eqref{eq:subscript_cond_X}. For massive partons, the point ${x_j=0}$ is cut off by the energy $\delta$-function in Eq.~\eqref{eq:rho_LFPG_ori}. For massless partons this point is singular, but measures $0$ after the integration. 
\subsection{Gibbs sampling}
\begin{figure*}
	\centering
	\includegraphics[width=\linewidth]{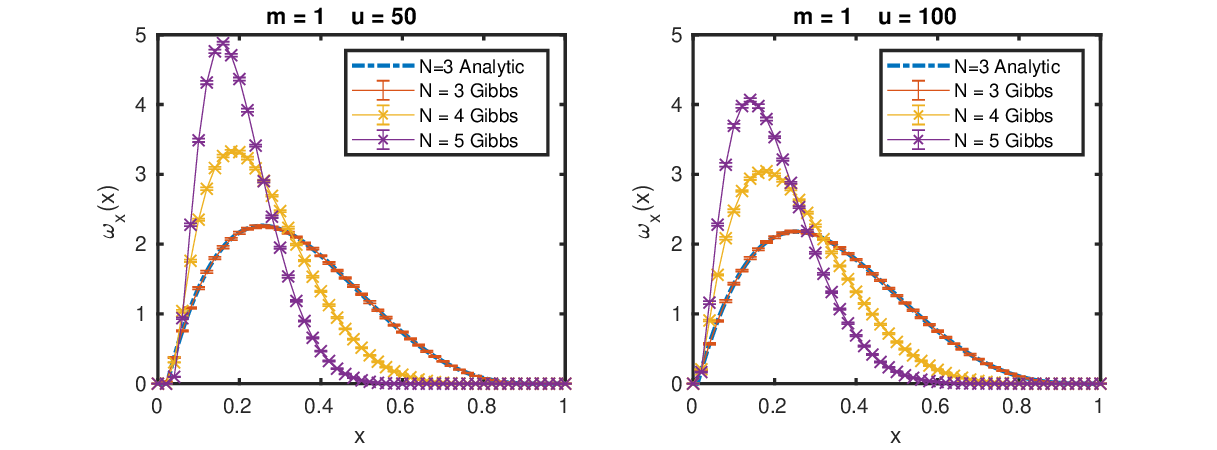}
	\caption{The single-particle longitudinal momentum-fraction distribution from light-front parton gas model with identical massive particles. The left panel shows results with $u=50\,m^2$. The right panel shows results with $u=100\,m^2$. The blue dot-dashed lines are the analytical results with $N=3$ using Eq.~\eqref{eq:omega_x_N3_massive}. The red, yellow, and purple points with error bars for the deviations correspond to the marginalization of the joint distribution in Eq.~\eqref{eq:omega_joint_x} using Gibbs sampling with $N=3$, $4$, and $5$ respectively. The means (points) and deviations (error bars) are calculated from $16$ independent samples for each combination of particle number $N$ and available thermal energy $u$.}
	\label{fig:gibbs}
\end{figure*}
\begin{figure*}
	\centering
	\includegraphics[width=\linewidth]{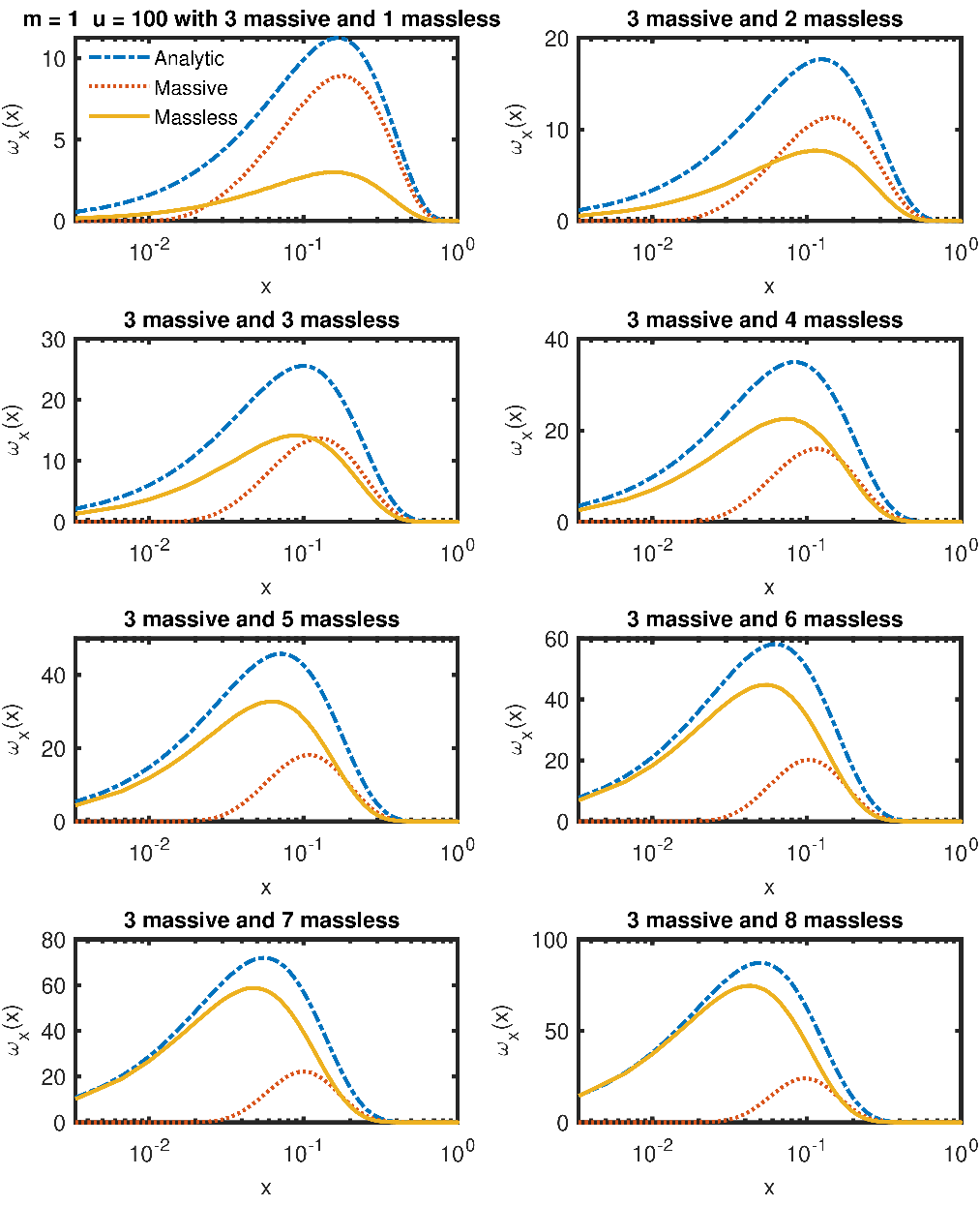}
	\caption{Longitudinal momentum-fraction distributions for mixtures of massless particles with $3$ massive ones when $u=100\,m^2$. Red dotted lines and yellow solid lines are obtained by marginalizing the joint distribution in Eq.~\eqref{eq:omega_joint_x} using Gibbs sampling respectively for massive and massless particles. Deviations are negligible compared to the curve thickness. Blue dot-dashed lines correspond to analytical results in Eq.~\eqref{eq:omega_x_N3_massive} when all particles are massless.}
	\label{fig:lfpg_4_2}
\end{figure*}

With more than $3$ massive partons, we do not have an analytical expression for the single-particle distribution $\omega_{x,i}(x_i)$. Instead we apply Markov Chain Monte Carlo algorithm to draw discrete samples of the joint distribution ${\omega(E,V,N,\mathbf{P};\mathbf{x})}$ given by Eq.~\eqref{eq:omega_joint_x}. Because the number of independent momentum fractions in Eq.~\eqref{eq:omega_joint_x} is $N-1$, each sample of the joint distribution is represented by a $(N-1)$-dimensional random vector $x_j^\lambda$. Here the subscript $j\in \{1,\,2,\,\dots,\,N-1 \}$ is the vector index. While the sample-point index is denoted by the superscript ${\lambda\in \{1,\,2,\,\dots,\,S\}}$, with $S$ being the sample size. 

We apply the multi-stage Gibbs sampling to obtain samples of the joint distribution in Eq.~\eqref{eq:omega_joint_x}, where one stochastic step only generates one component of the random vector~\cite{robert1999monte}. Specifically for each update of the variable $x^\lambda_j$, we apply the Metropolis--Hastings algorithm with the acceptance probability of
\begin{equation}
	\rho_{\mathrm{MH}}(x_j^\lambda,X_j^{\lambda+1})= \min \bigg\{ 1, \dfrac{ f(X_j^{\lambda+1}) \, q(x_j^{\lambda}\vert X_j^{\lambda+1}) }{ f(x_j^\lambda)\, q(X_j^{\lambda+1}\vert x_j^\lambda) } \bigg\},\label{eq:rho_MH}
\end{equation}
where the objective density $f(x)$ is specified by the conditional distribution 
\begin{align}
	& f( x_j^\lambda\vert x_1^{\lambda+1}, x_2^{\lambda+1}, \dots, x_{j-1}^{\lambda+1}, x_{j+1}^{\lambda}, \dots, x_{N-1}^{\lambda} ) \nonumber\\
	& \propto \omega( E,V,N, \mathbf{P}; x_1^{\lambda+1}, x_2^{\lambda+1}, \dots, x_{j-1}^{\lambda+1}, x_j^\lambda, x_{j+1}^{\lambda}, \dots, x_{N-1}^{\lambda}, x_N )
\end{align}
with $x_N=1-(x_1^{\lambda+1}+x_2^{\lambda+1}+\dots +x_{j-1}^{\lambda+1}+x_j^\lambda+x_{j+1}^\lambda+\dots +x_{N-1}^{\lambda} )$. The function ${\omega(E,V,N,\mathbf{P};\mathbf{x})}$ is given by Eq.~\eqref{eq:omega_joint_x} without the $\delta$-function. The instrumental probability distribution $q(x\vert y)$ is the normal distribution with zero mean and an adjustable variance $a$:
\begin{equation}
	q(x\vert y)=\dfrac{1}{\sqrt{2\pi a^2}}\exp\left[-\dfrac{(x-y)^2}{2a^2} \right].\label{eq:gaussian_instrumental}
\end{equation} 
Since the distribution $q(x\vert y)$ given by Eq.~\eqref{eq:gaussian_instrumental} is symmetric with respect to ${x\leftrightarrow y}$, Eq.~\eqref{eq:rho_MH} is simplified into
\begin{equation}
	\rho_{\mathrm{MH}}(x_j^\lambda,X_j^{\lambda+1})=\min \big\{ 1, \, f(X_j^{\lambda+1}) / f(x_j^\lambda)\big\}.
\end{equation}
We then assign ${x_j^{\lambda+1}=X_j^{\lambda+1}}$ with probability of ${\rho_{\mathrm{MH}}(x_j^\lambda,X_j^{\lambda+1})}$. Otherwise the relation ${x_j^{\lambda+1}=x_j^\lambda}$ is applied. Such a procedure is repeated from ${j=1}$ to ${j=N-1}$ in order to produce one sample point for the random vector $x_j$. 

For each sample, we choose ${a=0.1}$ in order to reach a sample size of $1\times10^6$ after discarding the first $1\times10^5$ sample points. The single-particle distribution $\omega_{x,j}(x_j)$ is obtained by directly binning and normalizing the sample with respect to a specific $x_j$. When particles are of the same mass, we then calculate the means and standard deviations of each bin in $\omega_x(x)$ as a histogram from $16$ independent samples. The results after normalization are shown in Fig.~\ref{fig:gibbs}. For $N=3$ they are in agreement with the analytical expression in Eq.~\eqref{eq:omega_x_N3_massive}. For a given thermal energy, the distribution is shifted toward smaller $x$ with larger $N$. This is understood as $x_j=1/N$ is the most probable fraction in the joint $x$-distribution when particles are of the same mass. The mean momentum fraction carried by a single parton is also $1/N$, which could be verified through ${\int_{0}^{1}dx\,x\,\omega_x(E,V,N,\mathbf{P},x)=1/N}$. At a fixed particle number $N$ the distribution becomes broader with higher thermal energy, as the support of the joint $x$-distribution increases. 

The same algorithm can also be used to draw discrete samples of the joint distribution ${\omega(E,V,N,\mathbf{P};\mathbf{x})}$ given by Eq.~\eqref{eq:omega_joint_x} when particles are not of the same mass. 
Specifically let us consider mixing a fixed number of identical massive particles with a selected number of massless particles. To illustrate the effect of flavor variation on PDFs of particles inside hadrons, a selected number of massless particles are added to $3$ massive ones. The numerical method previously utilized for systems of all identical particles is readily transferred, resulting in $2$ distinct single-particle $x$-distributions for this mixed system. The normalization of longitudinal momentum-fraction distributions in such a scenario is different from the case when all particles are of the same mass. Instead of integrating to unit probability, the sum of longitudinal momentum fractions carried by both types of particles weighted by their probability is $1$ due to momentum conservation. Integrations of each longitudinal momentum-fraction distributions are confirmed to be the corresponding particle numbers after such normalization. For a selection of $1$, $2$, $3$, \dots, $8$ massless particles, the resulting $x$-distributions when the available thermal energy is $100$ times the square of massive-particle mass are illustrated through $8$ panels in Fig~\ref{fig:lfpg_4_2}. The longitudinal momentum-fraction distributions of the massive particles and of the massless particles are respectively given by the red dotted curves and by the yellow solid curves. The analytical results when all particles become massless are shown as the blue dash-dot lines for comparison. The horizontal axis is chosen to be logarithmic in order to highlight the region where values of $x$ are small. As shown in this figure, this region is dominated by the distributions of the massless particles, due to the unavoidable energy penalty of massive particles when carrying small longitudinal momentum fractions. With an increase in the number of massless particles, the $x$-distributions of massless particles in this region approach those when all particles are massless. Adding massless particles also has the effect of shifting peaks of $x$-distributions for both types of particles towards small values of $x$, due to $1/N$ being the expected most probable fraction.

\begin{figure*}
	\centering
	\includegraphics[width=0.7\linewidth]{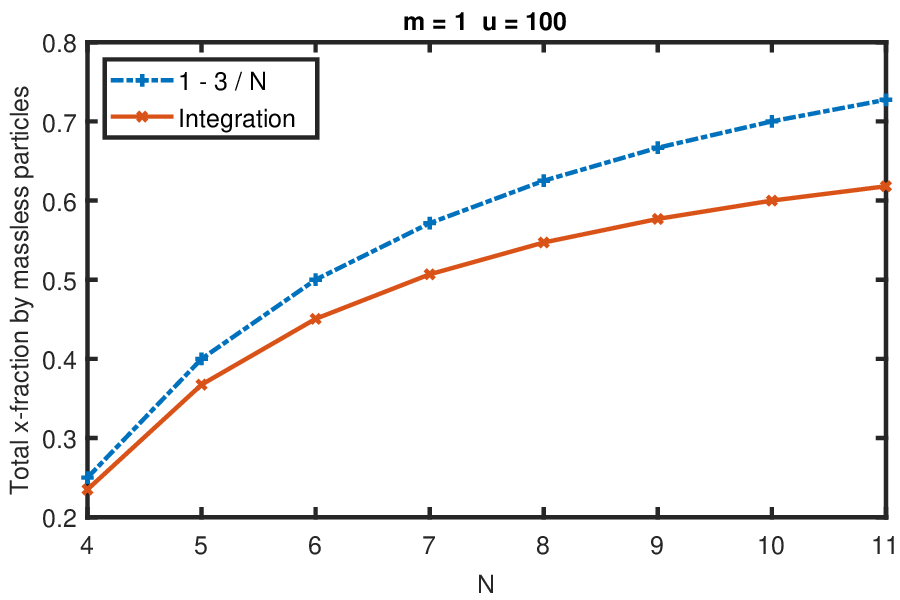}
	\caption{Total longitudinal momentum-fraction carried by massless particle when mixed with $3$ massive ones at thermal energy $u=100\,m^2$. The horizontal axis is the total number of particles. Points on the red solid line are results based on integrations of the $x$-distributions. The blue dash-dot line represents expected results when all particles are massless.}
	\label{fig:lfpg_x_quad}
\end{figure*}
\begin{table*}[h]
	\centering
	\begin{tabular}{c|cccccccc}
		\hline
		Particle number& $4$ & $5$ & $6$ & $7$ & $8$ & $9$ & $10$ & $11$ \\
		\hline
		Massless fraction & $0.235$ & $0.368$ & $0.451$ & $0.507$ & $0.547$ & $0.577$ & $0.600$ & $0.618$ \\
		\hline
	\end{tabular}
	\caption{Total longitudinal momentum fractions carried by massless particles when mixed with $3$ massive ones at available thermal energy $u = 100\, m^2$.}
	\label{tab:lfpg_x_quad}
\end{table*}
While the integrated total longitudinal momentum fraction is unity, the massive particles carry a disproportionate fraction. Based on the $x$-distributions obtained as in Fig.~\ref{fig:lfpg_4_2}, we calculate the longitudinal momentum factions carried by the added massless particles. The result is given in Table~\ref{tab:lfpg_x_quad} with a corresponding illustration in Fig.~\ref{fig:lfpg_x_quad}. Also shown in Fig.~\ref{fig:lfpg_x_quad} are the expected longitudinal momentum factions carried by the same set of particles when the massive particles become massless. Therefore the massive particles draw total longitudinal momentum of the system away from the massless ones. 
\section{Summary and conclusion\label{sc:summary}}
In this article we proposed the light-front parton gas model as a statistical method to help understand the distribution of the longitudinal momentum fractions of the partons inside a bound state based on light-front kinematics and momentum conservation. Specifically within such a model the partons were treated as classical particles whose phase-space distributions were given by the light-front generalization of the microcanonical molecular dynamics ensemble. 

We defined the longitudinal momentum-fraction distributions as the marginalization of the phase-space distribution indicated by the ensemble. We also derived the joint longitudinal momentum-fraction distribution through the analytical reduction of the transverse momentum integrals. This consequently resulted in the explicit expressions of the single-particle $x$-distributions for all massless partons and for ${N=3}$ massive partons of the same mass. With specific combinations of the parton number and available thermal energy, we calculated the single-particle momentum-fraction distribution using quadrature. Applying Gibbs sampling we drew samples of the joint longitudinal momentum-fraction distribution for massive partons. Based on these samples we computed the distribution ${\omega_x(E,V,N,\mathbf{P};x)}$ as histograms for total numbers $N=3$, $4$, and $5$ particles all of the same mass. The same sampling technique was subsequently used to compute longitudinal momentum-fraction distributions when $3$ massive particles of the same mass are mixed with a selection of massless particles. We found that the region with small longitudinal momentum fractions is occupied by the massless particles. Whereas the total longitudinal momentum of the system is distributed disproportionately to the massive particles.

This article serves as the initial investigation of a light-front parton gas model. Therefore despite the reproduction at certain theoretical limits and the overall qualitative agreement, we did not relate results presented in this article with experimental hadron PDFs. In future work we expect to expand the current model to include spin and flavor formalizing a mixed system of massive quarks and massless gluons. We also plan to allow creation and annihilation of partons that convert between quark-antiquark pairs and gluons, resulting in a microcanonical ensemble with a variable number of particles.

\section{Acknowledgments}
	Funding: This work was supported by the U.S. Department of Energy under Grant No. DE-SC0023692, Office of Science, Office of Nuclear Physics. S.\,J. was also supported by the U.S. Department of Energy, Office of Science, Office of Nuclear Physics, under Contract No. DE-AC02-06CH11357.
\appendix
\section{Hyperspherical angular measure\label{sc:angular_measure}}
For integrals in the $d$-dimensional Euclidean space, the integration measure can be written as a product of radial and angular measures:
\begin{equation}
\int d^{d}k=\int_{0}^{+\infty}dk~k^{d-1}\int d\Omega_d,
\end{equation}
where $k$ is the Euclidean norm of the $N$-vector $k$. While the integral $\int d\Omega_d$ covers the angular measure of the $d$-dimensional spherical coordinates. This $d$-dimensional angular measure corresponds to the Jacobian from the Cartesian coordinates to the spherical coordinates. Explicitly we parameterize the angular dependence of a vector in the $d$-dimensional Euclidean space by
\begin{subequations}\label{eq:spherical_coordinates}
	\begin{align}
	& x_1=x\sin\psi\left[\prod_{i=1}^{d-3}\sin\theta_i\right]\sin\phi, \\
	& x_2=x\sin\psi\left[\prod_{i=1}^{d-3}\sin\theta_i\right]\cos\phi, \\
	& \dots \nonumber \\
	& x_k=x\sin\psi\left[\prod_{i=k-1}^{d-3}\sin\theta_i\right]\cos\theta_{k-2} \quad (\mathrm{for} ~d-2 < k <d), \\
	& \dots \nonumber \\
	& x_d=x\cos\psi.
	\end{align}
\end{subequations}
Or more consistently define $\psi=\theta_{d-2}$ and $\phi=\theta_0$ such that
\begin{subequations}
	\begin{align}
	& x_1=x\left[\prod_{i=k-1}^{d-2}\sin\theta_i\right]\sin\theta_{0}, \\
	& x_k=x\left[\prod_{i=k-1}^{d-2}\sin\theta_i\right]\cos\theta_{k-2} \quad (\mathrm{for}~2\leq k\leq d).
	\end{align}
\end{subequations}
The Jacobian due to such a transformation is given by
\begin{equation}
	J_d = \mathrm{abs} \, \bigg\{ \dfrac{\partial(x_1,x_2,\dots,x_d)}{\partial(x,\psi,\theta_{d-3},\theta_{d-4},\dots,\theta_1,\phi)} \bigg\} = x^{d-1}\sin^{d-2}\psi\prod_{i=1}^{d-3}\sin^{i}\theta_i.
\end{equation}
The integral with respect to the angular measure of the hyperspherical coordinates in the $d$-dimension Euclidean space therefore becomes
\begin{equation}
	\int d\Omega_d = \int_{0}^{\pi}d\psi\sin^{d-2} \psi \, \bigg( \prod_{i=1}^{d-3}\int_{0}^{\pi}d\theta_i\sin^{i}\theta_i \bigg) \int_{0}^{2\pi}d\phi.
\end{equation}
Applying a proper orthogonal transformation we can find a coordinate system such that
\begin{equation}
	\sum_{j=1}^{d}k_j p_j=kp\cos\psi,
\end{equation}
which reduces the $\theta_i$ and $\phi$ angular integrals. We will see these variables contribute to the spherical measure of $d-1$ dimensions.

The following identity is useful when integrating with respect to the angular variables:
\begin{equation}
	I_n = \int_{0}^{\pi} \sin^n(\theta) d\theta = \dfrac{ \Gamma(1/2) \Gamma\left( (n+1)/2 \right) }{ \Gamma\left( (n+2)/2 \right) },\label{eq:def_I_n}
\end{equation}
where the Euler $\Gamma$-function is defined as
\begin{equation}
	\Gamma(s)=\int_{0}^{+\infty}dx \, x^{s-1}e^{-x},
\end{equation}
resulting in ${\Gamma(s+1)=s\Gamma(s)}$, ${\Gamma(1)=1}$, and ${\Gamma(1/2)}=\sqrt{\pi}$. After expressing $I_n$ as products of $\Gamma$-functions, we can easily calculate the $d$-dimensional spherical measure:
\begin{equation}
	\Omega_d=\int d\Omega_d = 2\pi \prod_{i=1}^{d-2}I_i = \dfrac{2[\Gamma(1/2)]^d}{\Gamma(d/2)}.\label{eq:Omega_d}
\end{equation}
Therefore when the only nontrivial angular dependence comes from $\psi$, the angular integrals simplify into 
\begin{equation}
	\int d\Omega_d=\Omega_{d-1}\int_{0}^{\pi}d\psi\sin^{d-2}\psi. \label{eq:reduced_spherical_measure}
\end{equation}
\section{Analytical integrals for the transverse momenta\label{sc:transverse_itg}}
In the process of calculating the joint longitudinal momentum-fraction distribution using the phase-space distribution of the light-front parton gas model with $N$ partons, one encounters the following integrals in the transverse momenta:
\begin{equation}
	T( \tilde{u}, N ) = \int d^{2N}l^\perp \delta \Big( \tilde{u} - \sum_{j=1}^{N} \overrightarrow{l}^{\perp 2}_j \Big) \delta \Big( \sum_{j=1}^{N} \xi_j \overrightarrow{l}^\perp _j \Big), \label{eq:def_T_N}
\end{equation}
where $\tilde{u}$ is recognized as the reduced available thermal energy is positive semi-definite. Since we have defined $\xi_j$ in relation to the longitudinal momentum-fractions by $\xi_j=\sqrt{x_j}$, the Euclidean norm of the vector $\xi_j$ is $1$ due to the conservation of the longitudinal momentum. 

The integral defined by Eq.~\eqref{eq:def_T_N} can be calculated exactly with the help of a geometric interpretation of the second $\delta$-function. Specifically there exists an orthogonal transformation $l_i'=\sum_{i=1}^{N}R_{ij}l_j$ such that 
\begin{equation}
	\sum_{j=1}^{N}\xi_j l_j=l'_N, 
\end{equation}
where $l_N'$ is the $N$-th component of the vector $l_i'$. Here we have used the property that the modulus of $\xi_j$ is $1$. Since $R_{ij}$ is an orthogonal matrix, we also have 
\begin{equation}
	\sum_{j=1}^{N}l_j^2=\sum_{j=1}^{N}l_j'^{\,2}.
\end{equation}
Meanwhile the Jacobian of such a transformation of the integral measure from $d^{N}l$ to $d^{N}l'$ is $1$. After relabeling $l'$ by $l$ Eq.~\eqref{eq:def_T_N} becomes
\begin{equation}
	T(\tilde{u},N) = \int d^N l ^x \int d^{N}l^y\, \delta \Big( \tilde{u} - \sum_{j=1}^{N} [ (l^{x}_j)^2 + (l^y_j)^2 ] \Big) \, \delta(l^x_N)\,\delta(l^y_N).
\end{equation}
We then introduce the spherical coordinates defined by Eq.~\eqref{eq:spherical_coordinates} such that 
\begin{align}
	& T(\tilde{u},N) = \int_{0}^{+\infty}(l^x)^{N-1}\,dl^x\, \int d\Omega^x_{N} \int_{0}^{+\infty}(l^y)^{N-1}dl^y\, \int d\Omega^y_{N} \, \delta ( \tilde{u}-(l^x)^2-(l^y)^2 ) \nonumber\\[1mm]
	& \times \, \delta(l^x\cos\psi^x)\,\delta(l^y\cos\psi^y) = \int_{0}^{+\infty}(l^x)^{N-1} dl^x\, \int_{0}^{+\infty}(l^y)^{N-1}dl^y \, \delta( \tilde{u}-(l^x)^2-(l^y)^2 ) \nonumber\\[1mm]
	& \times \Omega_{N-1} \int_{0}^{\pi}d\psi^x\, \sin^{N-2}\psi^x\, \delta(l^x\cos\psi^x) \, \Omega_{N-1} \int_{0}^{\pi}d\psi^y\, \sin^{N-2}\psi^y\, \delta(l^y\cos\psi^y),\label{eq:T_N_reduced}
\end{align}
where $l^x$ and $l^y$ are the Euclidean norms of $l_j^x$ and $l_j^y$, respectively. In deriving Eq.~\eqref{eq:T_N_reduced} we have applied Eq.~\eqref{eq:reduced_spherical_measure} to simplify the angular integrals. The remaining integrations with respect to $\psi^x$ and $\psi^y$ are identical in Eq.~\eqref{eq:T_N_reduced}. Specifically for these integrals we have 
\begin{equation}
	\int_{0}^{\pi}d\psi \, \sin^{N-2}\psi \, \delta(l\,\cos \psi) = 1 / \vert l \vert.
\end{equation}
Equation~\eqref{eq:T_N_reduced} them becomes
\begin{equation}
	T(\tilde{u},N) = \Omega_{N-1}^2 \int_{0}^{+\infty}(l^x)^{N-2}\,dl^x\,\int_{0}^{+\infty}(l^y)^{N-2}dl^y \, \delta( \tilde{u}-(l^x)^2-(l^y)^2 ).\label{eq:T_N_longi}
\end{equation}
To evaluate the longitudinal integrals in Eq.~\eqref{eq:T_N_longi}, we introduce the following polar coordinates for $(l^x,\,l^y)$:
\begin{equation}
	\begin{cases}
		l^x=l\cos\theta \\[1mm]
		l^y=l\sin\theta
	\end{cases}. 
\end{equation}
Notice that because both $l^x$ and $l^y$ are positive, the polar angle $\theta$ only goes from $0$ to $\pi/2$. Equation~\eqref{eq:T_N_longi} in such polar coordinates becomes 
\begin{align}
	& T(\tilde{u},N) = \Omega_{N-1}^2\int_{0}^{+\infty} dl\, l^{2N-3}\, \delta(\tilde{u}-l^2) \int_{0}^{\pi/2}d\theta \, \left( \cos\theta\, \sin\theta \right)^{N-2} = \Omega_{N-1}^2 \tilde{u}^{N-2}I_{N-2} / 2^N \nonumber\\
	& = \dfrac{ \pi ^{N-1/2}\, \tilde{u}^{N-2} }{ 2^{N-2}\,\Gamma\left( (N-1)/2 \right) \Gamma ( N/2 ) } ,\label{eq:trans_itg}
\end{align}
where we have applied Eqs.~\eqref{eq:def_I_n} and \eqref{eq:Omega_d}.
\bibliographystyle{elsarticle-num}
\bibliography{LFPG_bib}
\end{document}